\documentclass[12pt]{iopart}

\expandafter\let\csname equation*\endcsname\relax
\expandafter\let\csname endequation*\endcsname\relax 

\usepackage[english]{babel}
\usepackage{amsmath}
\usepackage{amssymb}
\usepackage{graphicx}
\usepackage{color}
\begin{document}

\title{Large deviations of the top eigenvalue of large Cauchy random matrices}

\author{Satya N. Majumdar, Gr\'egory Schehr, Dario Villamaina and Pierpaolo Vivo}

\address{Laboratoire de Physique Th\'eorique et Mod\`eles Statistiques, UMR 8626, Universit\'e Paris Sud 11 and CNRS, B\^at. 100, Orsay F-91405, France}
\ead{satya.majumdar@u-psud.fr,gregory.schehr@u-psud.fr,villamaina@gmail.com,\\pierpaolo.vivo@gmail.com}
\begin{abstract}
We compute analytically the probability density function (pdf) of the largest eigenvalue $\lambda_{\max}$
  in rotationally invariant Cauchy ensembles of $N\times N$ matrices. We consider unitary ($\beta = 2$), orthogonal ($\beta =1$) and symplectic ($\beta=4$) ensembles of such heavy-tailed random matrices. We show that a central non-Gaussian regime for $\lambda_{\max} \sim
  \mathcal{O}(N)$ is flanked by large deviation tails on both sides which we compute here exactly for any value of $\beta$. By matching these tails with the central regime, we obtain the exact leading asymptotic behaviors of the pdf in the central regime, which generalizes the Tracy-Widom distribution known for Gaussian ensembles, both at small and large arguments and for any $\beta$. Our analytical results are confirmed by numerical
  simulations.
\end{abstract}

%\pacs{02.50.Ga, 05.40.Fb, 05.45.Tp}
\maketitle

%\section{Introduction}
Characterizing the distribution of extreme eigenvalues of random
matrices is of paramount interest in many applications. A cornerstone in this field is the discovery of the Tracy-Widom
(TW) laws \cite{tracywidom} for the \emph{typical} fluctuations of the largest
eigenvalue of Gaussian random matrices. Let us first recall a few basic definitions and results for the Gaussian ensembles. Consider a $N\times N$ matrix $\mathbf{X}$ whose upper-triangular entries are drawn at random from a standard Gaussian distribution in the real $(\beta=1)$, complex $(\beta=2)$ or quaternion $(\beta=4)$ domain. Symmetrizing, we end up with a symmetric, Hermitian or quaternionic self-dual random matrix (respectively) whose $N$ eigenvalues $\lambda_1, \ldots, \lambda_N$ are real random variables, characterized by the joint probability distribution function~(jpdf):
\begin{equation}
P_{\rm joint}(\lambda_1,\ldots,\lambda_N)=B_N(\beta) e^{-\frac{\beta}{2}\sum_{i=1}^N\lambda_i^2}\prod_{j<k}|\lambda_j-\lambda_k|^\beta \;, \label{jpdgauss}
\end{equation}
where the normalization constant $B_N(\beta)$ is known from the celebrated Selberg's 
integral. This joint law (\ref{jpdgauss}) allows one to interpret the $\lambda_i$'s 
as the positions of charged particles repelling each other via a $2d$-Coulomb 
(logarithmic) potential: they are confined on a $1d$ line and each one is subject to 
a harmonic potential. Here we focus on the largest eigenvalue 
$\lambda_{\mathrm{max}} = {\max}_{1\leq i \leq N} \lambda_i$. What can we say about 
its statistical properties? This is a non-trivial question because, due to the 
all-to-all interaction term $\prod_{j<k}|\lambda_j-\lambda_k|^{\beta}$, the standard 
results 
about extreme value statistics of \emph{independent} random variables~\cite{gumbel} 
(namely, the existence of just three universality classes Gumbel, Fr\'echet or 
Weibull) are no longer applicable.

The \emph{average} location of the largest eigenvalue is readily determined by the shape of the average density of the eigenvalues in the large $N$ limit. For a Gaussian random matrix of large size $N$, the average density of eigenvalues (normalized to unity) $\rho_N(\lambda)=\langle\frac{1}{N}\sum_i\delta (\lambda-\lambda_i)\rangle$ has a semi-circular shape on the compact support $[-\sqrt{2N},\sqrt{2N}]$ called the Wigner semicircle
\begin{equation}\label{wigner}
\rho_N(\lambda)\approx \frac{1}{\sqrt{N}}\tilde \rho_W\left(\frac{\lambda}{\sqrt{N}}\right)\qquad\mbox{ with }\tilde \rho_W(x)=\frac{1}{\pi}\sqrt{2-x^2} \;.
\end{equation}
It thus follows that the average location of the largest eigenvalue is given for large $N$ by the upper edge of the density support:
\begin{equation}
\langle{\lambda_{\mathrm{max}}}\rangle\approx \sqrt{2N} \;.
\end{equation}
However, the largest eigenvalue fluctuates from one realization of the matrix to another. What can be said about the \emph{full probability density} of $\lambda_{\mathrm{max}}$?

From the jpdf \eqref{jpdgauss} it is easy to write the cumulative distribution of $\lambda_{\mathrm{max}}$ as a multiple integral:
\begin{equation}
\mathbb{P}(\lambda_{\mathrm{max}}\leq w)=B_N(\beta)\prod_{i=1}^N \int_{-\infty}^w d\lambda_i \; {P_{\rm joint}}(\lambda_1,\ldots,\lambda_N) \;,
\end{equation}
which can be interpreted as the partition function of a Coulomb gas in the presence of a hard wall in $w$. Carrying out this multiple integration is a non-trivial task. It turns out that it is necessary to deal separately with \emph{two} natural scales for the fluctuations in the asymptotic limit. \emph{Typical} (small) fluctuations scale as $\sim N^{-1/6}$, while \emph{atypical} (large) fluctuations scale as $N^{1/2}$, and the two corresponding distributions are described by different functional forms:
\begin{itemize}
\item {\bf Typical fluctuations:} Forrester \cite{forr} followed by Tracy and Widom \cite{tracywidom} realized that in the large $N$ limit, the largest eigenvalue follows the law 
\begin{equation}
\lambda_{\mathrm{max}}\approx \sqrt{2N}+ a_\beta N^{-1/6}\chi_\beta
\end{equation}
with $a_{1,2} = 1/\sqrt{2}$ and $a_4 = 2^{-7/6}$ and where the random variable $\chi_\beta$ has an $N$-independent distribution, $\mathbb{P}(\chi_\beta\leq x)={\cal F}_\beta(x)$ called the Tracy-Widom distribution \cite{tracywidom}, which has highly asymmetric tails:
%
%For example, for $\beta=2$ we have
%\begin{equation}
%F_2(x)=\exp\left[-\int_x^\infty (z-x)q^2(z)dz\right]\label{F2}
%\end{equation}
%where $q(z)$ satisfies a Painlev\'e II equation
%\begin{equation}
%q^{\prime\prime}=2 q^3+z q
%\end{equation}
%with the boundary condition $q(z)\sim\mathrm{Ai}(z)$ as $z\to\infty$, where $\mathrm{Ai}(z)$ is the Airy function. Note that the density of the largest eigenvalue, obtained by differentiation of \eqref{F2}
%has highly asymmetric tails,
\begin{align}
{\cal F}_\beta^\prime (x) &\sim \exp\left[-\frac{\beta}{24}|x|^3\right]\qquad\mbox{ as }x\to -\infty \;,\\
&\sim\exp\left[-\frac{2\beta}{3}x^{3/2}\right]\qquad\mbox{ as }x\to\infty \;.
\end{align}

These TW distributions also describe the top eigenvalue statistics of large real
\cite{john,soscov} and complex \cite{johann} Gaussian Wishart matrices, which play an important role in Principal Component Analysis of large datasets. Amazingly, the same TW distributions have emerged in a
number of a priori unrelated problems \cite{maj} such as the longest increasing
subsequence of random permutations \cite{baik}, directed polymers \cite{johann,poli} and growth models \cite{growth}
in the Kardar-Parisi-Zhang universality class in 1+1 dimensions, sequence alignment problems \cite{sequence},
mesoscopic fluctuations in quantum dots \cite{dots}, height fluctuations of non-intersecting Brownian motions \cite{FMS11,Lie12} and also in finance
\cite{biroli}. Remarkably, the TW distributions associated to the Gaussian Unitary and Orthogonal Ensembles have been
recently observed in experiments on nematic liquid crystals \cite{takeuchi}.

\item {\bf Atypical fluctuations:} the TW laws do not account for \emph{atypically} large
fluctuations, e.g. of order $\mathcal{O}(\sqrt{N})$ around the mean value $\sqrt{2N}$. Questions related to large deviations of extreme eigenvalues have recently emerged in cosmology \cite{aazami} and disordered systems
\cite{satyadean1,satyadean2}, and in the assessment of the efficiency of data compression \cite{majverg}. Recently, the large deviations of the largest eigenvalue
of Wishart matrices have been measured in experiments involving coupled fiber lasers \cite{davidson}. 
The probability of atypical large fluctuations, to leading order for large $N$, is
usually 
described by two large deviation (or rate) functions 
$\psi_{-}(x)$ (for fluctuations to the \emph{left} of the mean) and $\psi_{+}(x)$ (for fluctuations to the \emph{right} of the mean), in such a way that the probability density of the largest eigenvalue reads:
\begin{equation}\label{regimes_gaussian}
\frac{d}{dw} \mathbb{P}(\lambda_{\max} \leq w) \approx
\begin{cases}
\exp\left[-\beta N^2 \psi_{-}\left(\frac{w}{\sqrt{N}}\right)\right]&,\,  w<\sqrt{2N} \, \& \, |w-\sqrt{2N}|\approx\mathcal{O}(\sqrt{N})\\
\frac{1}{a_\beta N^{-1/6}}{\cal F}_\beta^\prime\left(\frac{w-\sqrt{2N}}{a_\beta N^{-1/6}}\right)&, \, \hspace*{2.1cm} |w-\sqrt{2N}|\approx\mathcal{O}(N^{-1/6})\\
\exp\left[-\beta N \psi_{+}\left(\frac{w}{\sqrt{N}}\right)\right]&, \, w>\sqrt{2N} \, \& \, |w-\sqrt{2N}|\approx\mathcal{O}(\sqrt{N}) \;.
\end{cases}
\end{equation}
%The precise meaning of the symbol $\approx$ is as follows:
%\begin{align}
%\lim_{N\to\infty} \frac{1}{\beta N^2}\ln \mathcal{P}(\lambda_{\mathrm{max}} &=z\sqrt{N})=-\psi_{-}(z) \qquad\mbox{ for } z<\sqrt{2}\\
%\lim_{N\to\infty} \frac{1}{\beta N}\ln \mathcal{P}(\lambda_{\mathrm{max}} &=z\sqrt{N})=-\psi_{+}(z) \qquad\mbox{ for } z>\sqrt{2}
%\end{align}
Note that while the TW distribution ${\cal F}_\beta (x)$, describing
the central part of the probability distribution of $\lambda_{\rm max}$, depends 
explicitly on $\beta$, the
two leading order rate functions $\psi_{\mp}(z)$ are independent of $\beta$.
Exploiting a simple   
physical method based on a Coulomb gas analogy (see below), 
the left rate function $\psi_{-}(z)$ was first explicitly computed
in~\cite{satyadean1,satyadean2}, while the right rate function $\psi_{+}(z)$
was computed in~\cite{majverg}. A more complicated, 
albeit mathematically rigorous derivation (but only valid for $\beta=1$) of 
$\psi_{+}(z)$ in 
the context of spin 
glass models can be found in \cite{ben}. 
More recently, the various subleading corrections to the leading behavior
have been explicitly computed using more sophisticated methods both for the left 
tail~\cite{BEMN}, as well
as for the right tail~\cite{NM1,BN1}. Note also that large deviations for the smallest eigenvalue were also studied for Wishart \cite{katzav_1} and Jacobi \cite{katzav_2} ensembles. 

The physical mechanism 
responsible for the left tail is very different from the one on the right. 
Computing the probability distribution $\mathbb{P}(\lambda_{\max} \leq w)$
of the top eigenvalue is equivalent to computing the free energy of
an interacting Coulomb gas in presence of a hard wall at 
$w$~\cite{satyadean1,satyadean2}. 
For the  
left tail of $\mathbb{P}(\lambda_{\max} \leq w)$, with $w < \sqrt{2N}$, the charge density is {\it 
pushed} by the wall, 
which leads to a complete reorganization of all the $N$ charges and thus costs an 
energy difference, 
compared to the Wigner sea (\ref{wigner}), of order ${\cal O}(N^2)$.
In contrast, for the 
right tail, the dominant fluctuations are caused by {\it pulling} a single charge 
away 
from this Wigner sea and the energy difference is only of order ${\cal 
O}(N)$~\cite{majverg}. The central TW region of the
distribution in Eq. (\ref{regimes_gaussian}), describing typical
fluctuations,  matches smoothly
with the two tail behaviors at its flanks~\cite{satyadean1,satyadean2,majverg}.

\end{itemize}

Given the apparent robustness of the predictions stemming from random matrices with 
Gaussian independent entries, it is of paramount interest to investigate whether 
these TW distributions and their large deviation tails for $\lambda_{\max}$ actually 
hold for a broader class of matrices. Results are available for $1)$ non-invariant 
ensembles, with independent and identically distributed (i.i.d.) entries, and $2)$ invariant ensembles, and are summarized 
here:

\begin{enumerate}
\item[1)] It is known that TW holds asymptotically for symmetric $N\times N$ matrices with i.i.d. entries of variance
$1/N$, such that all moments are finite \cite{sos}. On the other hand, when the
distribution of i.i.d. entries decays as a power law, $\sim
|M_{ij}|^{-1-\mu}$, the case $\mu=4$ leads to a new class of limiting
distribution, while for $\mu>4$ the TW still holds asymptotically and
for $\mu<4$ the statistics of the largest eigenvalue is governed by a Fr\'echet law \cite{biroli,soshnikov_fyo,ruzmaikina,benarous}. 
%{\bf [cite Soshnikov \& Fyodorov, and perhaps expand with other references. For example, check the paper Order Statistics and Ginibre’s Ensembles 
%by B. Rider, and also A. Soshnikov. Universality at the edge of the spectrum in Wigner matrices. 
%Comm. Math. Phys. 207, pp. 697–733 (1999).]}
\item[2)] Much fewer results are known for rotationally invariant ensembles. The TW law has been established for the classical Wishart ensembles in \cite{john,soscov,johann} and for the Jacobi ensemble in \cite{johnstonjacobi}. Disordered ensembles of random matrices were studied in  \cite{bohigas} where continuous transitions between TW and other extreme value distributions were found. In the context of L\'evy stable ensembles, the largest eigenvalue distribution for the so-called L\'evy-Smirnov
ensemble has been derived in \cite{wiec}.

%{\bf [check Universality of a family of Random Matrix Ensembles with logarithmic 
%soft-conﬁnement potentials by Jinmyung Choi and K.A. Muttalib ]}: nothing about \lambda_\max in that paper
\end{enumerate}

Here we address these challenging questions on top eigenvalue statistics beyond TW and
focus on yet another instance of random matrices which incorporates (i) highly non-Gaussian statistics and (ii) correlated 
entries, while retaining rotational invariance. We consider the Cauchy ensemble of $N\times N$ matrix $\mathbf{H}$, which might be symmetric ($\beta = 1$), Hermitian ($\beta=2$) or quaternionic ($\beta = 4$). The weight associated to $\mathbf{H}$ is given by
%
%, which assigns to a $N\times N$ matrix $\mathbf{H}$, which might be symmetric ($\beta = 1$), Hermitian ($\beta=2$) or quaternionic ($\beta = 4$), a %weight given by:
\begin{equation}\label{PH}
  P(\mathbf{H})\propto\left[\det (\mathbf{1}_N+\mathbf{H}^2)\right]^{-\beta (N-1)/2 -1} \;,
\end{equation}
where $\mathbf{1}_N$ denotes the $N \times N$ identity matrix. Note that Eq.~(\ref{PH}) is manifestly invariant under the similarity
transformation $\mathbf{H}\to \mathbf{U}\mathbf{H}\mathbf{U}^{-1}$,
with $\mathbf{U}$ an orthogonal $(\beta=1)$, unitary $(\beta=2)$ or
symplectic $(\beta=4)$ matrix. Due to this invariance, the jpdf of the $N$ real eigenvalues can be straightforwardly written
as:
\begin{equation}\label{Cauchy_distr}
P_{\rm joint}(\lambda_1,\ldots,\lambda_N)\propto \prod_{i=1}^N \frac{1}{(1+\lambda_i^2)^{\beta (N-1)/2+1}}\prod_{j<k}|\lambda_j-\lambda_k|^\beta \;.
\end{equation}
As in the Gaussian case, this expression (\ref{Cauchy_distr}) allows to interpret the $\lambda_i$'s as the positions of charged particles (with say positive unit charge) repelling each other via the $2d$-Coulomb (logarithmic) interaction. Here they are confined on the real line and interact in addition with a fixed particle, with charge $-(N-1+2/\beta)$, which is placed at the point of coordinate $(0,1)$ in the complex plane. 

This ensemble has been studied in the literature in Ref.~\cite{brouwer} in the context of mesoscopic transport as a model of a quantum dot which is coupled to the outside world by non ideal leads containing $N$ scattering channels. It is also one of the paradigmatic examples of application of free probability theory in the context of random matrices \cite{burda,burdarev}. A remarkable difference with the standard Gaussian ensemble is that the density $\rho_N(\lambda)$ is independent of $N$, $\rho_N(\lambda)  = \rho^*(\lambda)$ (for all values of $N$), and its support is unbounded and given, for any $\beta$, by \cite{tierz_stable,forrester_book}
%The Cauchy ensemble exhibits a few peculiar features, namely:
%\begin{itemize}
%\item If $\mathbf{H}$ is distributed according to \eqref{PH}, then
%  $I)$ $\mathbf{H}^{-1}$ is also distributed according to \eqref{PH},
%  and $II)$ every $n\times n$ submatrix of $\mathbf{H}$ obtained by
%  erasing $N-n$ rows and columns is distributed according to
 % \eqref{PH} with $N$ replaced by $n$. These properties have been
%  crucial in establishing the Poisson kernel law in the context of
%  mesoscopic transport in non-ideal quantum dots \cite{brouwer}.
%\item The orthogonal polynomials with respect to the Cauchy weight are
%  Jacobi polynomials analytically continued to complex arguments
%  [...]. In contrast to the classical cases, only a \emph{finite}
% number of orthogonal polynomials do exist for this ensemble.
%\item A peculiarity of this model is that the average density of eigenvalues is given by:
\begin{equation}
  \rho^*(\lambda)=\frac{1}{\pi}\frac{1}{1+\lambda^2} \;, \; \lambda \in (-\infty,\infty)\label{density} \;.
\end{equation}
%exactly \emph{for any} $N$ {\color{red} ALSO FOR $\beta \neq 2$???}
%(and not just asymptotically for large $N$). {\bf [Comment on the fact that it is interesting to compute the larg. eig. dist. for an ensemble with power-law density?]}
%\end{itemize}

In this paper, we study the statistics of the largest eigenvalue $\lambda_{\max}$ for the Cauchy ensemble in Eq. (\ref{Cauchy_distr}) for any $\beta$, while up to now only the case $\beta = 2$ was considered in Ref. \cite{witte_forrester2000,najnudel2009} -- and even in this case the large deviations of $\lambda_{\max}$ have not been computed. 
%
%give a complete characterization of the largest
%eigenvalue for the Cauchy ensemble of heavy-tailed and unitarily
%invariant random matrices. In particular, we find the analogue of the Tracy-Widom
%distribution for typical (small) fluctuations around the mean as well as the large deviation tails on both sides.
%
To estimate the typical scale of $\lambda_{\mathrm{max}}$, denoted by ${\Lambda_{\rm max}}$, 
we note that for a general matrix model it satisfies
\begin{equation}\label{gen_EVS}
\int_{{\Lambda_{\mathrm{max}}}}^\infty \rho_N(\lambda) d \lambda \approx 1/N \;,
\end{equation}
as the fraction of eigenvalues to the right of the maximum
(including itself) is typically $1/N$. Substituting $\rho_N(\lambda) = \rho^*(\lambda)$, from (\ref{density}), in (\ref{gen_EVS}), one obtains that $\Lambda_{\mathrm{max}} \sim \mathcal{O}(N)$. As we show below, this gives rise to three distinct regimes in the fluctuations of $\lambda_{\max}$ [as in Eq. (\ref{regimes_gaussian}) for Gaussian ensembles, where in that case $\Lambda_{\max} \sim {\cal O}(\sqrt{N})$]. 

\vskip 0.3cm

{\noindent {\bf Summary of new results:}} Before presenting the details of the 
calculations, it is
  useful first to summarize our main results. 
We show that the pdf of $\lambda_{\max}$, $P(w,N) = 
\partial_w \mathbb{P}(\lambda_{\max} \leq w)$, displays three different regimes
to leading order for large $N$
\begin{equation}
  P(w,N)\approx\left\{\begin{array}{cc}\exp[-\beta N^2 \psi (w)] &  
w\ll N\\ N^{-1}f_\beta(w/N) & w\sim N \\N \phi(w) & w\gg N\end{array}\right. 
\label{main_result}
\end{equation}
where $\psi(w)$ and $\phi(w)$ are independent of $\beta$ and are given by
\begin{eqnarray}\label{expr_psi}
\psi(w)= \frac{1}{4} \left(\frac{1}{2} \log
    \left(w^2+1\right)-\log{\left(w + \sqrt{w^2+1} \right)}\right)+\frac{\log (2)}{4} \;,
\end{eqnarray}
and
\begin{eqnarray}\label{expr_phi}
\phi(w) = \frac{1}{\pi w^2} \;.
\end{eqnarray}
The full expression of the scaling function $f_\beta(x)$, describing
typical fluctuations and playing the 
role analogous to the TW 
distributions in the Gaussian case (\ref{regimes_gaussian}), 
can be computed explicitly only for $\beta=2$ 
\cite{witte_forrester2000,najnudel2009}, where it can 
be expressed in terms of the solution of a Painlev\'e-V equation 
[see Eqs (\ref{f2x}), (\ref{eq_tau}) below]. The function $f_2(x)$
has its support over $[0,\infty)$. From the Painlev\'e-V equation, we can check 
explicitly that
$f_2(x)$  has the following 
asymptotic
tails: $f_2(x)\sim \exp\left(-1/{8x^2}\right)$ as $x\to 0$ and
$f_2(x)\sim 1/{\pi x^2}$ as $x\to \infty$. While we were unable
to compute $f_{\beta}(x)$ analytically for arbitrary $\beta$, we are however able
to predict its precise tails for any $\beta>0$
\begin{eqnarray}
f_{\beta}(x) \sim 
\begin{cases}\label{asympt}
&\exp{\left(-\dfrac{\beta}{16x^2} \right)} \;, \; x \to 0 \\
&\dfrac{1}{\pi x^2} \;, \hspace*{1.7cm} x \to \infty \;.
\end{cases}
\label{tails_beta}
\end{eqnarray}
We achieve this by smoothly matching the central regime $\lambda_{\max} \sim 
{\cal O}(N)$ with the left ($\lambda_{\max} \ll N$) and the right
($\lambda_{\max} \gg N$)
large deviation tails that we can compute for arbitrary $\beta$. 

By comparing the results for the Gaussian case in Eq. (\ref{regimes_gaussian}) and the 
Cauchy 
case in Eq. (\ref{main_result}), we see that the behavior for the right tail
are rather different in the two cases. Indeed, quite generally, one can show (see 
later) 
that far to the right of the central peak, the pdf $P(w,N)$ can be simply
expressed in terms of the average density of states as $P(w,N)\approx N 
\rho_N(\lambda)$. In the Gaussian case, $\rho_N(\lambda)$ has a bounded
support in the $N\to \infty$ limit (Wigner semi-circle). However, for
finite but large $N$, there is a nontrivial correction to the density
arising from the vicinity of the edge of the semi-circle~\cite{forrester_higherorder}, 
which leads to
the nontrivial right large deviation behavior described by $\psi_+(z)$ in Eq.~(\ref{regimes_gaussian}). In contrast, in the Cauchy case, the average density
of states has an unbounded support and there does not seem to be
any nontrivial edge corrections for finite $N$ as in the Gaussian case.
 
\vskip 0.3cm

{\noindent {\bf Right large deviation tail:}} Let us start by deriving the 
expression 
of the pdf $P(w,N)$ 
for the right tail, i.e. where $w \gg N$. For this purpose, we define, for each 
eigenvalue $\lambda_i$, a binary variable $\sigma_i$ such that $\sigma_i = 1$ if $\lambda_i \geq w$ and $\sigma_i=0$ if $\lambda_i <w$. Then, the probability that the region $[w,\infty)$ is free
of eigenvalues can be written as
\begin{equation}\label{prod_sigma}
\mathbb P(\lambda_{\max} \leq w) = \langle [1-\sigma_1][1-\sigma_2]\cdots[1-\sigma_N]\rangle \;,
\end{equation}
where the average $\langle\cdot\rangle$ is over the jpdf of the eigenvalues (\ref{Cauchy_distr}). Expanding the product in (\ref{prod_sigma}), one gets
\begin{equation}\label{prod_sigma_exp}
\mathbb P(\lambda_{\max} \leq w) = 1- N \int_w^{\infty} \rho_N(\lambda) d\lambda + \mbox{two-point} + \mbox{three-point} +\ldots \;,
\end{equation}
where '$\mbox{two-point}$' means a double integral involving two-point correlation function (and similarly for  '$\mbox{three-point}$' etc). When $w\to \infty$, keeping $N$ fixed, which corresponds to  the extreme right tail, one can show that all higher order contributions vanish \cite{forrester_higherorder}. Hence, substituting $\rho_N(\lambda) = \rho^*(\lambda)$, from (\ref{density}), in (\ref{prod_sigma_exp}) one obtains 
\begin{equation}\label{expr_lefttail}
\mathbb P(\lambda_{\max} \leq w) \approx 1- N \int_w^{\infty} \rho^*(\lambda) d\lambda \;.
\end{equation}
Taking the derivative of Eq. (\ref{expr_lefttail}) with respect to (w.r.t.) $w$, and using $\rho^*(\lambda) \sim 1/(\pi \lambda^2)$ for $\lambda \gg 1$ (\ref{density}), yield the result announced in Eq. (\ref{main_result}) for $w \gg N$ and $\phi(w) = 1/(\pi w^2)$~(\ref{expr_phi}).

\vskip 0.3cm

\begin{figure}
\begin{center}
\includegraphics[width=0.50\linewidth]{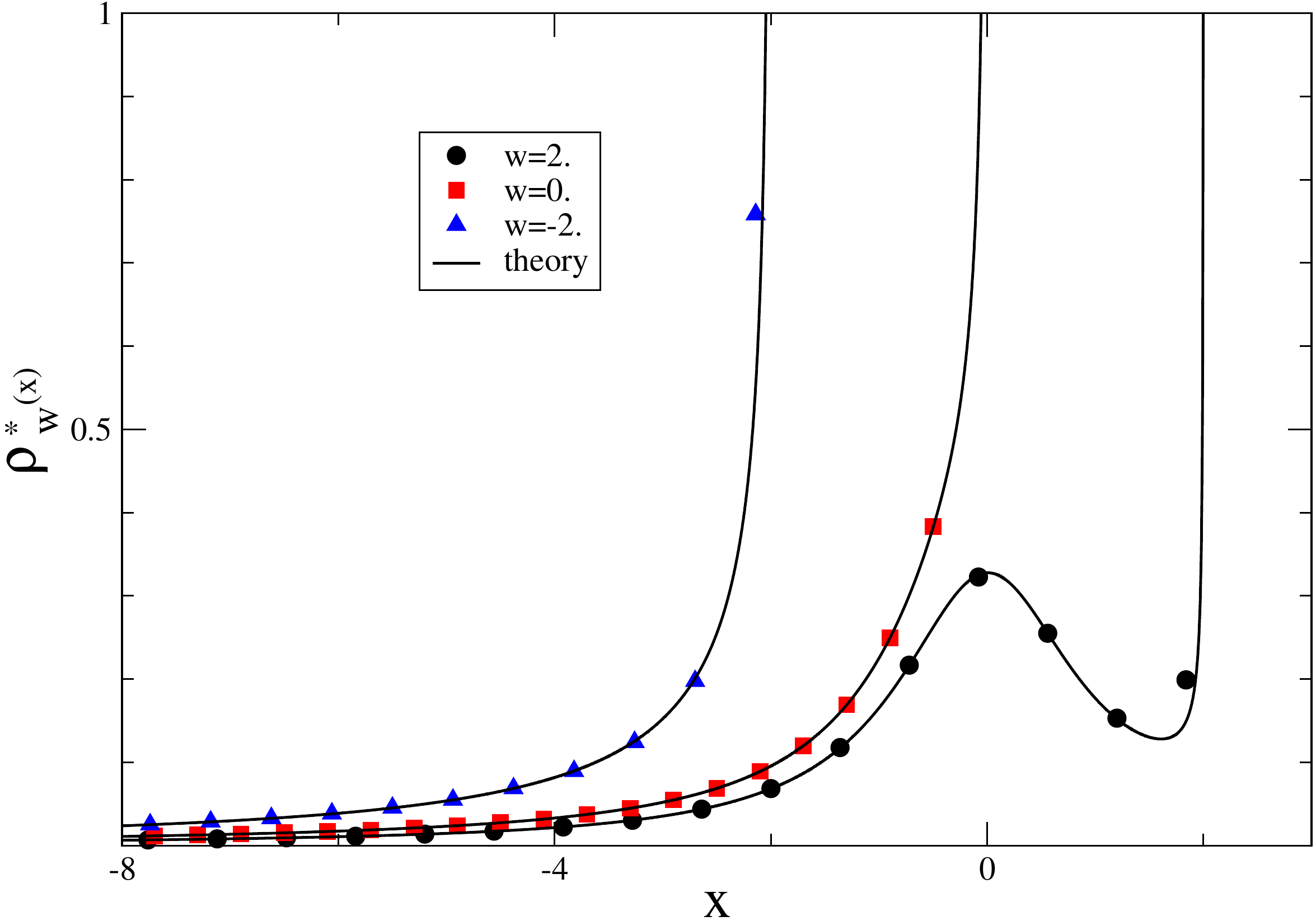}
\caption{Plot of $\varrho_w^\star(x) = \hat\varrho_w(w-x)$ in Eq. (\ref{rhotau}) for different values of $w$ (solid lines), compared to the results of our numerical simulations (symbols), for which $\beta = 2$ and $N=100$.}\label{fig_density}
\end{center}
\end{figure}

{\noindent {\bf Left large deviation tail:}} Let us now focus on the left tail $w\ll 
N$, which we tackle using a 
Coulomb gas technique. By definition:
\begin{equation}
\mathbb{P}[\lambda_{\mathrm{max}} \leq w]=\frac{\int_{(-\infty,w]^N}\prod_{i=1}^N d\lambda_i\ P_{\rm joint}(\lambda_1,\ldots,\lambda_N)}{\int_{(-\infty,\infty)^N}\prod_{i=1}^N d\lambda_i\ P_{\rm joint}(\lambda_1,\ldots,\lambda_N)}\label{defint} \;.
\end{equation}
We can represent $P_{\rm joint}(\lambda_1,\ldots,\lambda_N)$ in the Boltzmann form, $P_{\rm joint}(\lambda_1,\ldots,\lambda_N)\propto\exp(-(\beta/2) E[\{\mbox{\boldmath$\lambda$}\}])$
where the energy function is given by:
\begin{equation} \label{energy_coulomb}
E[\{\mbox{\boldmath$\lambda$}\}]=\left(N-1+\frac{2}{\beta}\right)\sum_{i=1}^N\ln(1+\lambda_i^2)-\sum_{i\neq j}\ln|\lambda_i-\lambda_j| \;.
\end{equation}
This thermodynamical analogy, originally due to Dyson \cite{dyson}, allows to
treat the system of eigenvalues as a $2d$ gas of charged particles (Coulomb
gas) confined on the real line, in equilibrium under competing interactions.  In the large $N$
limit, the Coulomb gas with $N$ discrete charges becomes a continuous
gas which can be described by a continuum (normalized to unity)
density function $\varrho_w (x)=(1/N)\sum_i \delta
(x-\lambda_i)$. Consequently, one can replace the original multiple
integral in \eqref{defint} by a functional integral over the space of
$\varrho_w (x)$. This procedure, originally introduced by Dyson \cite{dyson}
for the problem without a wall, was first used successfully for 
the Gaussian case with a wall in \cite{satyadean1,satyadean2} and
subsequently was found useful in a  
number of different contexts
(see for example \cite{vivojpa,vivoother1,vivoother2} and references therein). 
Introducing the constrained
density of eigenvalues $\varrho_w(x)$, one has
%
%and using the identity
%\begin{equation}
%\sum_{i=1}^N \varphi(\lambda_i) = N\int d\lambda \varrho_w(\lambda) \varphi(\lambda) \;,
%\end{equation}
%valid for any smooth $\varphi$ in the large $N$ limit, we can write for large $N$:
\begin{equation}
\mathbb{P}[\lambda_{\mathrm{max}} \leq w]
\propto\int\mathcal{D}[\varrho_w] 
\exp\left(-\frac{\beta N^2}{2}\mathcal{S}_w[\varrho_w]\right) \;,
\label{functional}
\end{equation}
where the action $\mathcal{S}_w[\varrho_w]$ is given by:
%\begin{widetext}
\begin{eqnarray}
\mathcal{S}_w[\varrho_w]=&&\int_{-\infty}^w \ln (1+x^2)\varrho_w(x)dx-\int_{-\infty}^w\int_{-\infty}^w dx\ dx^\prime \varrho_w(x)\varrho_w(x^\prime)\ln|x-x^\prime| \nonumber \\ 
&& +C\left(\int_{-\infty}^w dx \varrho_w(x)-1\right)\label{action} \;,
\end{eqnarray}
%\end{widetext}
where $C$ is a Lagrange multiplier enforcing the normalization of the density to $1$. 
The cumulative distribution in Eq. (\ref{functional}) must be normalized, i.e.,
the proportionality constant in Eq. (\ref{functional}) is determined 
from the condition: $\mathbb{P}[\lambda_{\mathrm{max}} \leq w]\to 1$ as $w\to 
\infty$. 
The physical meaning of the action $\mathcal{S}_w[\varrho_w]$ is readily understood as 
the free energy of the Coulomb gas, whose particles are constrained to lie to the left of a hard wall at $x=w$. The fact that the free energy is proportional to $N^2$ (and not just $N$)
is a consequence of the strong all-to-all interactions among the particles: the free energy of this correlated gas is dominated by the energetic component $\sim\mathcal{O}(N^2)$,
while the entropic term [$\sim \mathcal{O}(N)$] is subdominant in the large $N$ limit (recently
this entropic term has been computed explicitly for the Gaussian~\cite{allez1}
and the Wishart-Laguerre ensemble~\cite{allez2}). 
For large $N$, the functional integral in Eq. (\ref{functional}) can be evaluated
by the saddle point method.
Clearly, the equilibrium configuration of the gas (the saddle point density) in the 
presence of the hard wall at $x=w$ will be determined by
minimizing the action in \eqref{action}. Following the result in the Gaussian 
case~\cite{satyadean1,satyadean2}, we can anticipate that the modified 
equilibrium (saddle point) density $\varrho_w^\star(x)$ in presence of the wall will 
display an integrable divergence as $x\to w^{-}$.
In terms of the equilibrium density $\varrho_w^\star(x)$, the probability of the largest eigenvalue reads from (\ref{functional}):
\begin{equation}
\mathbb{P}[\lambda_{\mathrm{max}} \leq w]\approx \exp\left[-\beta N^2\psi(w)\right] \;, 
\; \psi(w)=\frac{1}{2}\left(\mathcal{S}_w[\varrho_w^\star]-
\mathcal{S}_\infty[\varrho_\infty^\star] \right) \;. 
\end{equation}
The additional term $\mathcal{S}_\infty[\varrho_\infty^\star]$ comes
from the normalization constant.
In order to compute $\varrho_w^\star(x)$, we vary the action $\frac{\delta\mathcal{S}}{\delta \varrho_w(x)}=0$, yielding:
\begin{equation}
\ln(1+x^2)+C=2\int_{-\infty}^w dx^\prime \varrho_w^\star(x^\prime)\ln |x-x^\prime| \;, \; x \in (-\infty,w]\label{Tricomilog} \;.
\end{equation}
Differentiating (\ref{Tricomilog}) w.r.t. $x$ and setting $x^\prime=w-\tau$ and $x=w-z$ yields:
%\begin{comment}
%\begin{equation}
%\mathrm{Pr}\int_{-\infty}^w dx^\prime\frac{\varrho_w^\star(x^\prime)}{x-x^\prime}=\frac{x}{1+x^2}\label{Tricomi}
%\end{equation}
%\end{comment}
\begin{equation}
\mathrm{Pr}\int_0^\infty d\tau \frac{\hat\varrho_w(\tau)}{\tau-z}=\frac{w-z}{1+(w-z)^2}\label{halfhilbert} \;,
\end{equation}
where $\mathrm{Pr}$ stands for Cauchy's principal part and the shifted density $\hat\varrho_w(\tau)=\varrho_w^\star(w-\tau)$ is introduced. 
%
%\begin{comment}
%In terms of the shifted density $\hat\varrho_w(\tau)$, the action reads \cite{supp}:
%\begin{widetext}
%\begin{equation}
%\mathcal{S}_w[\hat\varrho_w]=\frac{1}{2}\int_{0}^\infty \ln \left(1+(w-\tau)^2\right)\hat\varrho_w(\tau)d\tau-\int_0^\infty d\tau\ln|w-\tau|\hat\varrho_w(\tau)\label{actionSnew}
%\end{equation}
%\end{widetext}
%
%\end{comment}
The inversion of the half-Hilbert transform \eqref{halfhilbert} is the
main technical challenge. However, the task can be fully accomplished
\cite{tricomi,paveri}, and the general solution of the equation (\ref{halfhilbert}) of the Tricomi type can be written as:
\begin{equation}\label{start_expr_rho}
\hat\varrho_w(\tau)=-\frac{1}{\pi^2 \sqrt{\tau}}\left\{{\mathrm{Pr}\int_0^\infty ds \frac{1}{s-\tau}\frac{\sqrt{s}(w-s)}{1+(w-s)^2}}+B\right\} \;, \; \tau \geq 0 \:,
\end{equation}
where $B$ is a constant enforcing the normalization $\int_0^\infty d\tau \hat\varrho_w(\tau)=1$. After straightforward manipulations one finds that $B=0$ and eventually 
\begin{equation} \label{rhotau}
\hat\varrho_w(\tau)= \frac{1}{\pi\sqrt{2\tau}}
\begin{cases}
\tfrac{(k_{w}+w)^{1/2}-(w-\tau)(k_{w}-w)^{1/2}}{1+(w-\tau)^2} & w \geq 0  \;, \\
\tfrac{((\tau w-k^{2}_{w})(w-\tau)+k_{w}(w-\tau)^{2}+k_w)(k_{w}+w)^{1/2}+\tau(w-\tau)(k_{w}-w)^{1/2}}{k_{w}[(w-\tau)^{2}+1]}& w\leq 0 \;,
\end{cases}
\end{equation}
%\end{widetext}
where $k_{w}=\sqrt{w^{2}+1}$. 
One can check that $\int_0^\infty d\tau \hat\varrho_w(\tau)=1$ 
and $\hat\varrho_w(w-\tau)\sim \frac{1}{\pi (1+\tau^2)}$ for $w\to\infty$, 
which yields the result announced above (\ref{density}). In Fig. \ref{fig_density}, we show a plot of $\varrho_w^\star(x)$ for different values of $w$, and compare our exact analytical expression to numerical simulations (see below for more details about them).

\vskip 0.3cm

{\noindent {\it Evaluation of the saddle point action:}} 
From the solution $\varrho_w^\star(x)$ (depending parametrically on $w$) we can evaluate the action (\ref{action})
as:
\begin{equation}
\mathcal{S}_w[\varrho_w^\star]=
\int_{-\infty}^w \ln (1+x^2)\varrho_w^\star(x)dx-
\int_{-\infty}^w\int_{-\infty}^w dx\ dx^\prime \varrho_w^\star(x)
\varrho_w^\star(x^\prime)\ln|x-x^\prime|
 \;.
\label{actionstar}
\end{equation}
The double integral in (\ref{actionstar}) can be written in terms of a single integral, using the following trick. We multiply (\ref{Tricomilog}) by $\varrho_w^\star(x)$
and integrate over $x$ to get eventually:
\begin{equation}
\int_{-\infty}^w\int_{-\infty}^w dx\ dx^\prime \varrho_w^\star(x)\varrho_w^\star(x^\prime)\ln|x-x^\prime|=\frac{1}{2}\int_{-\infty}^w dx \varrho_w^\star (x)\ln (1+x^2)+\frac{C}{2} \;.
\end{equation}
Therefore, the action reads:
\begin{equation}\label{start_expr}
\mathcal{S}_w[\varrho_w^\star]=\frac{1}{2}\int_{-\infty}^w \ln (1+x^2)\varrho_w^\star(x)dx-\frac{C}{2} \;.
\end{equation}
Given the different expression of $\hat\varrho_w(\tau)$ for $w\geq 0$ and $w\leq 0$ (\ref{rhotau}), these two cases have to be treated separately. We first consider the case $w > 0$ where the constant $C$ can be determined by setting $x=0$ in (\ref{Tricomilog}), yielding:
\begin{equation}
C=2\int_{-\infty}^w dx\ln|x|\varrho_w^\star(x) \;.
\end{equation}
Thus, eventually:
\begin{equation}
\mathcal{S}_w[\varrho_w^\star]=\frac{1}{2}\int_{-\infty}^w \ln (1+x^2)\varrho_w^\star(x)dx-\int_{-\infty}^w dx\ln|x|\varrho_w^\star(x)\label{actionS} \;.
\end{equation}
\vspace{20pt}
Let us focus on the action $\mathcal{S}_w[\varrho_w^\star]$ in (\ref{actionS}). Setting $x=w-\tau$ and recalling that $\hat\varrho_w(\tau)=\varrho_w^\star(w-\tau)$, we can rewrite the action, for $w>0$, as:
\begin{equation}
{\mathcal{S}_w[\hat\varrho_w]=\frac{1}{2}\int_{0}^\infty \ln \left(1+(w-\tau)^2\right)\hat\varrho_w(\tau)d\tau-\int_0^\infty d\tau\ln|w-\tau|\hat\varrho_w(\tau)}\label{actionSnew} \;,
\end{equation}
where $\hat \varrho_w(\tau)$ is given in Eq. (\ref{rhotau}). 
%Noting that the shifted density is the difference of two contributions, in total the action consists of four integrals to be computed, $\mathcal{S}_w[\hat%\varrho_w]=I_1+I_2-I_3-I_4$
%given by:
%\begin{align}
%I_1 &=\frac{1}{2\pi\sqrt{2}}\left(w+\sqrt{1+w^2}\right)^{1/2}\int_0^\infty d\tau\frac{\ln\left(1+(w-\tau)^2\right)}{\sqrt{\tau}\ \left[1+(w-\tau)^2\right]}\\
%I_2 &=\frac{(\sqrt{1+w^2}-w)^{1/2}}{\pi\sqrt{2}}\int_0^\infty\frac{d\tau}{\sqrt{\tau}}\frac{(w-\tau)\ln|w-\tau|}{\left[1+(w-\tau)^2\right]}\\
%I_3 &=\frac{1}{2\pi\sqrt{2}}\left(\sqrt{1+w^2}-w\right)^{1/2}\int_0^\infty d\tau\frac{\ln\left(1+(w-\tau)^2\right)(w-\tau)}{\sqrt{\tau}\ \left[1+(w-\tau)^2\right]}\\
%I_4 &=\frac{(\sqrt{1+w^2}+w)^{1/2}}{\pi\sqrt{2}}\int_0^\infty\frac{d\tau}{\sqrt{\tau}}\frac{\ln|w-\tau|}{\left[1+(w-\tau)^2\right]}
%\end{align}
After cumbersome manipulations, these integrals can be computed explicitly and we finally obtain a remarkably simple expression:

\begin{equation}\label{action_saddle}
{\mathcal S}_w[\hat\varrho_w] = \frac{1}{2} \left(\frac{1}{2} \log \left(w^2+1\right)-\log{\left(w +\sqrt{w^2+1} \right)}\right)+\frac{3 \log (2)}{2} \;.
\end{equation}
In particular, it admits the large $w$ expansion
\begin{eqnarray}\label{asympt_rate}
{\mathcal S}_w[\hat\varrho_w] = \log{2} + \frac{1}{8w^2} + {\cal O}(w^{-4}) \;.
\end{eqnarray}

For $w<0$, we start again with the expression in Eq. (\ref{start_expr}) but in this case $C$ can not be evaluated by setting $x=0$ in Eq. (\ref{Tricomilog}) 
as $x=0$ is not in the support on $\varrho_w^\star(x)$ [see Eq.~(\ref{Tricomilog})]. One can however evaluate $C$ by setting $x=w$ in Eq. (\ref{Tricomilog}) to get
\begin{eqnarray}\label{c_neg}
C = 2 \int_{-\infty}^w \varrho^\star_w(x)\log|w-x| \, dx - \log{(1+w^2)} \;.
\end{eqnarray}
After some manipulations, one can finally show that, for $w<0$, the expression of ${\mathcal S}_w[\hat \varrho_w]$ is also given by Eq. (\ref{action_saddle}). Therefore, as announced in Eq.~(\ref{main_result}), the right tail of the cumulative distribution of $\lambda_{\rm max}$ is given by
\begin{align}
\mathbb{P}[\lambda_{\mathrm{max}} \leq w] &\approx \exp\left(-\beta N^2\psi(w)\right)   \nonumber  \\
  \psi(w) &=\frac{1}{2}\left(\mathcal{S}_w[\hat\varrho_w]-\mathcal{S}_\infty[\hat\varrho_w]\right) = \frac{1}{4} \left(\frac{1}{2} \log
    \left(w^2+1\right)-\log{\left(w + \sqrt{w^2+1} \right)}\right) \nonumber \\
    &+\frac{\log (2)}{4} \;, \label{large_dev}
\end{align}
The rate function $\psi(w)$
has the following asymptotic tails
\begin{eqnarray}
\psi(w) & \approx & 1/[16 w^2]\quad {\rm as}\,\, w\to +\infty \;, \label{psi_right}\\
& \approx & \frac{1}{2}\ln \left(|w|/2\right) + {\cal O}(|w|^{-2}) \quad {\rm as}\,\, w\to 
-\infty \;. \label{psi_left} 
\end{eqnarray}
Substituting the asymptotic tail of $\psi(w)$ as $w\to -\infty$ in 
$\mathbb{P}[\lambda_{\mathrm{max}} \leq w] \approx \exp\left(-\beta 
N^2\psi(w)\right)$ then predicts a power law tail
of the distribution as $w\to -\infty$:
$\mathbb{P}[\lambda_{\mathrm{max}} \leq w] \sim [2/|w|]^{\beta N^2/2}$.
 
For $\beta=2$, it is interesting to compare the result (\ref{large_dev}) with 
previous works on the pdf of $\lambda_{\max}$, $P(w,N)$ \cite{witte_forrester2000,najnudel2009}. 
In \cite{witte_forrester2000}, the authors derived a nonlinear differential equation for the function
\begin{eqnarray}\label{def_sigma}
\sigma(w,N) = (1+w^2) \frac{P(w,N)}{\int_{-\infty}^w P(z,N) \, dz}=
(1+w^2) \frac{\partial }{\partial w} \ln \mathbb{P}[\lambda_{\mathrm{max}} \leq w] 
\;.
\end{eqnarray}
They found that, for arbitrary positive $N$, $\sigma(w,N)\equiv \sigma$ satisfies a 
nonlinear differential equation 
\begin{eqnarray}\label{exact_equation_wf}
&&(1+w^2)^2 (\sigma'')^2 + 4 (1+w^2)(\sigma')^3 - 8 w \sigma (\sigma')^2 + 
4 \sigma^2 \sigma' + 4 N^2 (\sigma')^2 = 0 \;,
\end{eqnarray}
with the asymptotic property
\begin{eqnarray}\label{boundary_condition}
\sigma(w,N) \to 0 \;, w \to \infty \;.
\end{eqnarray}
Using the results of Ref. \cite{CS93}, it was shown in Ref.~\cite{witte_forrester2000} that Eq. 
(\ref{exact_equation_wf}) can be transformed into a Painlev\'e-VI equation.

Let us now check if our result in Eq. (\ref{large_dev}), after setting $\beta=2$,
is compatible with the above differential equation in the large $N$ limit.
Let us first evaluate $\sigma(w,N)$ defined in Eq. (\ref{def_sigma}) by using
our large $N$ prediction in Eq. (\ref{large_dev}). Evidently, for large $N$ but 
finite $w$, we anticipate that $\sigma(w)$ must be proportional to $N^2$, i.e.,
\begin{eqnarray}
\sigma(w,N) \to N^2 \hat \sigma(w) \;,
\end{eqnarray}
where the function $\hat \sigma(w)$ is independent of $N$ and our result predicts that 
\begin{eqnarray}\label{our_result_sigma}
\hat \sigma(w) = - (1+w^2) \psi'(w) = \frac{1}{2} (\sqrt{1+w^2}-w) \;.
\end{eqnarray}
On the other hand, substituting $\sigma(w,N) = N^2 \hat \sigma(w)$ in
the differential equation (\ref{exact_equation_wf})
and keeping only the leading order ${\mathcal O}(N^6)$ terms, one finds that
$\hat \sigma(w)$ must satisfy the following nonlinear differential equation
\begin{eqnarray}\label{equation_largeN}
(1+w^2) (\hat \sigma')^2 - 2 w \hat \sigma \hat \sigma' + \hat \sigma^2 + \hat \sigma' 
= 0 \;.
\end{eqnarray}
Hence, in order to be compatible, our predicted form $\hat \sigma(w)=\frac{1}{2} 
(\sqrt{1+w^2}-w)$ must satisfy this nonlinear differential equation
(\ref{equation_largeN}). Indeed, one can check quite easily
that $\hat \sigma(w)=\frac{1}{2}
(\sqrt{1+w^2}-w)$ does satisfy Eq. (\ref{equation_largeN}),
with the correct 
boundary condition (\ref{boundary_condition}): this thus provides a non-trivial 
check of 
our result for the rate function $\psi(w)$ (\ref{large_dev}). 

\vskip 0.3cm

{\noindent {\bf The central part:}}
We now come to the central part of the pdf of $\lambda_{\max}$ where $\lambda_{\max} \sim {\cal O}(N)$. From standard scaling arguments one expects, for $w \sim {\cal O}(N)$
\begin{eqnarray}\label{scaling}
P(w,N) = \frac{1}{N} f_\beta\left(\frac{w}{N}\right) \;,
\end{eqnarray}
for any value of $\beta$, as announced in (\ref{main_result}). For $\beta=2$, one can check that it is consistent with the large $N$ limit of (\ref{exact_equation_wf}) from which one obtains~\cite{najnudel2009}
\begin{equation}
f_2(x) = \frac{d}{dx} \exp{\left[ \int_{0}^{1/x} \tau(z) \frac{dz}{z} \right]} \label{f2x} \;, \; x > 0 \;,
\end{equation}
where $\tau(x)\equiv \tau$ satisfies a Painlev\'e V equation on the interval $[0, +\infty)$
\begin{equation}\label{eq_tau}
x^2(\tau'')^2 + 4 (\tau-x \tau')\left(\tau - \tau'(x+\tau')\right) = 0 \;.
\end{equation}
Interestingly, the same equation (\ref{eq_tau}) also appears in the expression of the pdf of the (scaled) spacing between consecutive eigenvalues  in the bulk of the spectrum of GUE random matrices~\cite{JMMS80}. Hence one expects that $f_2(x)$ can also be expressed in terms of a Fredholm determinant involving a kernel very similar to the sine-kernel~\cite{forrester_book,mehta_book}. This can be shown directly from Eqs. (\ref{Cauchy_distr}) and (\ref{defint}), using orthogonal polynomials together with the asymptotic analysis performed in Ref. \cite{borodin_olshanski} to obtain \cite{najnudel2009}
\begin{eqnarray}\label{fd_expr}
f_2(x) = \partial_x \det(\mathbb{I} - \Pi_x K \Pi_x) \;, \; K(y_1,y_2) = \frac{1}{\pi} \frac{\sin(1/y_1 - 1/y_2)}{y_1-y_2} \;,
\end{eqnarray}    
where $\mathbb{I}$ denotes the identity operator and the operator $\Pi_x$ is the projector on the interval $[x, + \infty)$. In Eq. (\ref{fd_expr}), it is easy to see that under the change of variable $y  \to 1/(\pi y)$, $K(y_1,y_2)$ translates into the standard sine-kernel \cite{mehta_book,forrester_book}. 

What are the asymptotic behaviors of $f_2(x)$ ? From Eq. (\ref{eq_tau}), it is easy to see that $\tau(x) \sim - x^2/4$, as $x \to \infty$. This yields the small argument behavior of $f_2(x)$, from (\ref{f2x}): 
\begin{eqnarray}\label{f2_small}
\ln f_2(x) \sim -\frac{1}{8x^2} \;, \; x \to 0 \;.
\end{eqnarray}
On the other hand, from Eq. (\ref{eq_tau}), one obtains the small argument behavior of $\tau(x) \sim - \lambda x$, when $x \to 0$, but the constant $\lambda$ remains unspecified by this equation. Its determination can however be obtained by analyzing the Fredholm determinant in Eq. (\ref{fd_expr}) which yields $\tau(x) \sim -x/\pi$, for $x \to 0$  \cite{JMMS80}, and finally
\begin{eqnarray}\label{f2_large}
f_2(x) \sim \frac{1}{\pi x^2} \;, \; x \to \infty \;.
\end{eqnarray}

As we show now, these asymptotic behaviors for $\beta = 2$ in Eqs. (\ref{f2_small}) and (\ref{f2_large}) can be obtained straightforwardly, and generalized to any value of $\beta$, by exploiting the matching between the central part and the right and left tails of the distribution of $\lambda_{\max}$. Indeed, if we study the distribution of $\lambda_{\max}$ when $\lambda_{\max}$ approaches its typical value $\Lambda_{\max} \sim {\cal O}(N)$ from below, i.e. from the left tail, one expects
from our result above (\ref{large_dev}) that, given that $\psi(w) \sim 1/(16w^2)$ for $w \gg 1$
\begin{eqnarray}\label{largearg_left}
\mathbb{P}(\lambda_{\max} \leq w) \sim \exp{\left[-N^2 \frac{\beta}{16w^2} \right]} \;, w \gg 1 \; \& \; w \ll N \;.
\end{eqnarray} 
This expression (\ref{largearg_left}) is obviously a function of the scaled variable $x=w/N$ and coincides with the small argument behavior of the pdf $f_\beta(x)$ when $x=w/N \ll 1$. This yields the asymptotic behavior announced in the introduction (\ref{asympt}) and coincides, in particular, with the above result for $\beta=2$ (\ref{f2_small}). 

Similarly, if one studies the pdf of $\lambda_{\max}$ when $\lambda_{\max}$ 
approaches its typical value $\Lambda_{\max} \sim {\cal O}(N)$ from above, i.e. from the right tail, one expects
from our result in (\ref{main_result}) that the pdf behaves like
\begin{eqnarray}
P(w,N) \sim \frac{N}{\pi w^2} \;, \; w \gg N \;,
\end{eqnarray}
which is obviously a function of the scaled variable $w/N$ times $1/N$. If one assumes that this behavior matches with 
the large argument behavior of the central part where $P(w,N)\sim  (1/N) f_\beta(w/N)$ where $w \gg N$, one obtains that $f_\beta(x) \sim 1/(\pi x^2)$ for large $x$, as announced in (\ref{main_result}).  

%%% RISCRIVERE DA QUI %%%%
%We now show how the main peak in the central part, namely in the regime $w\sim N$, is governed by a simple scaling relation.\\
%  \\
%  \\
%  \vspace{1cm}
\begin{figure}
\begin{center}
\includegraphics[totalheight=0.20\textheight]{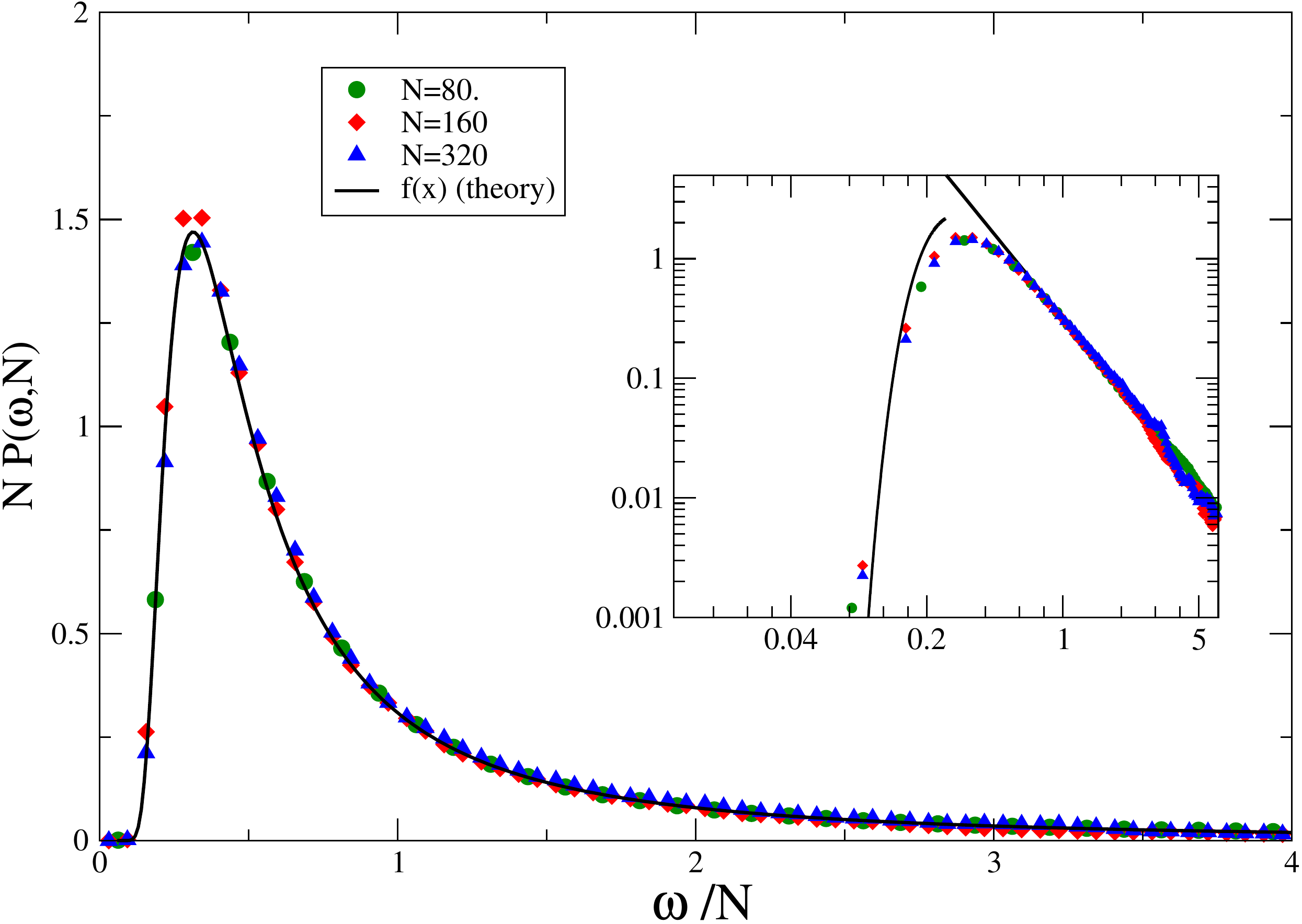}
\caption{The scaling relation $P(w,N)=N^{-1}f_2(w/N)$ for $\beta = 2$ is tested against numerical simulations varying $N$. Montecarlo data are compared
  with the solution of Eq.~(\ref{eq_tau}) showing perfect agreement. Inset:
  the asymptotic behavior of the tails (\ref{asympt}) is verified.}\label{fig1}
\end{center}
\end{figure}   

\vskip 0.3cm

\begin{figure}
\begin{center}
\includegraphics[width=0.33\linewidth]{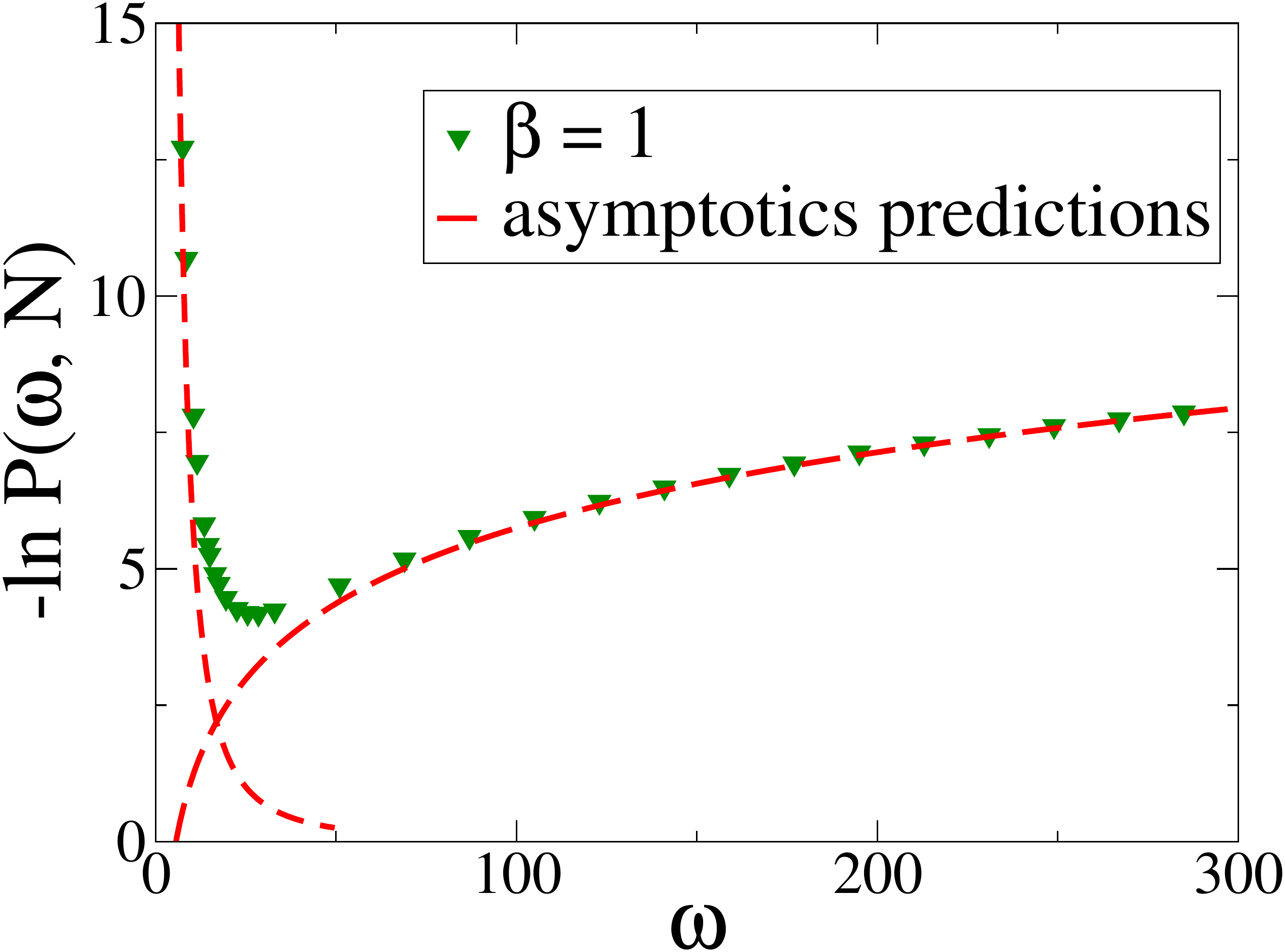}\includegraphics[width=0.33\linewidth]{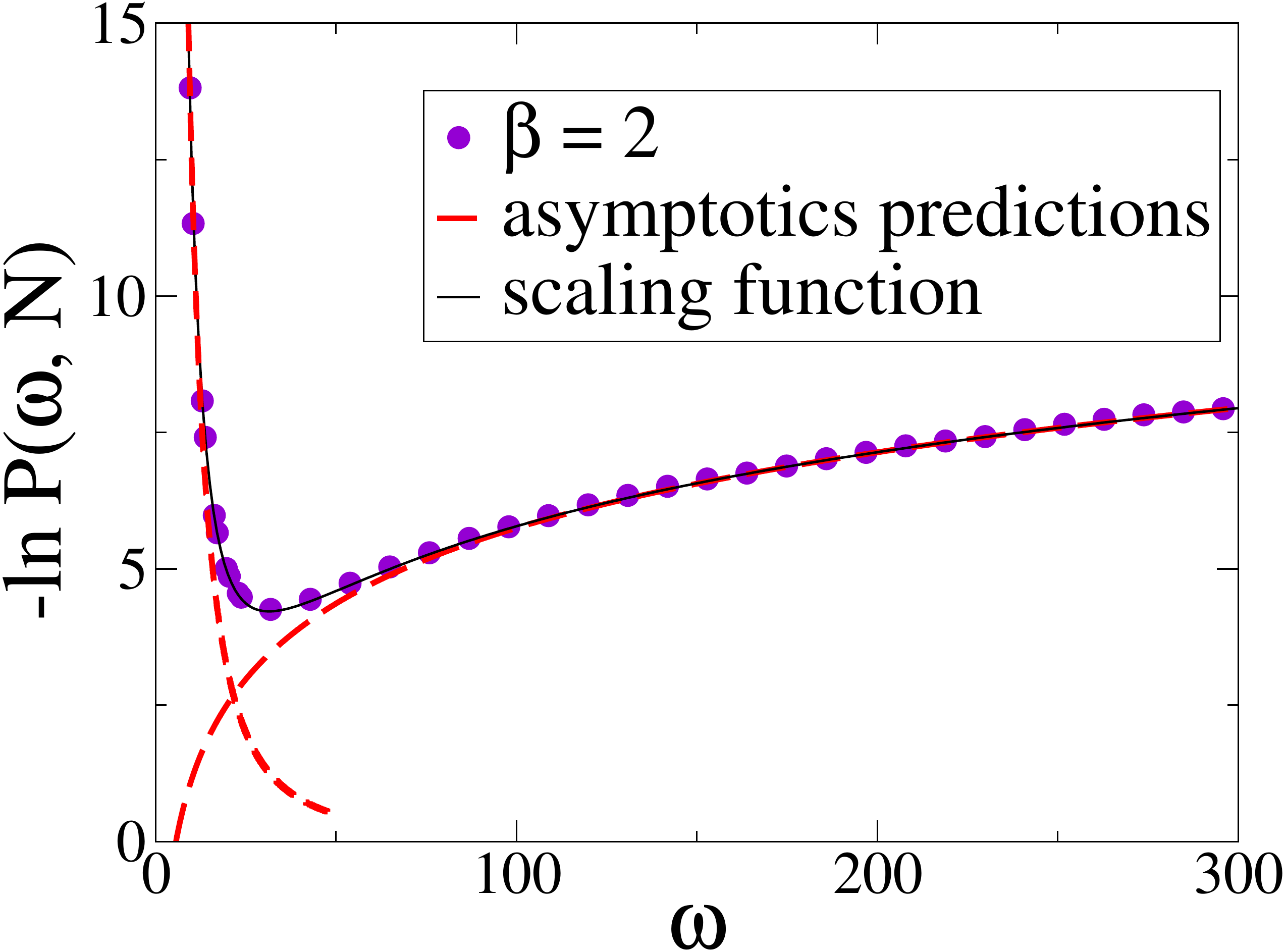}\includegraphics[width=0.33\linewidth]{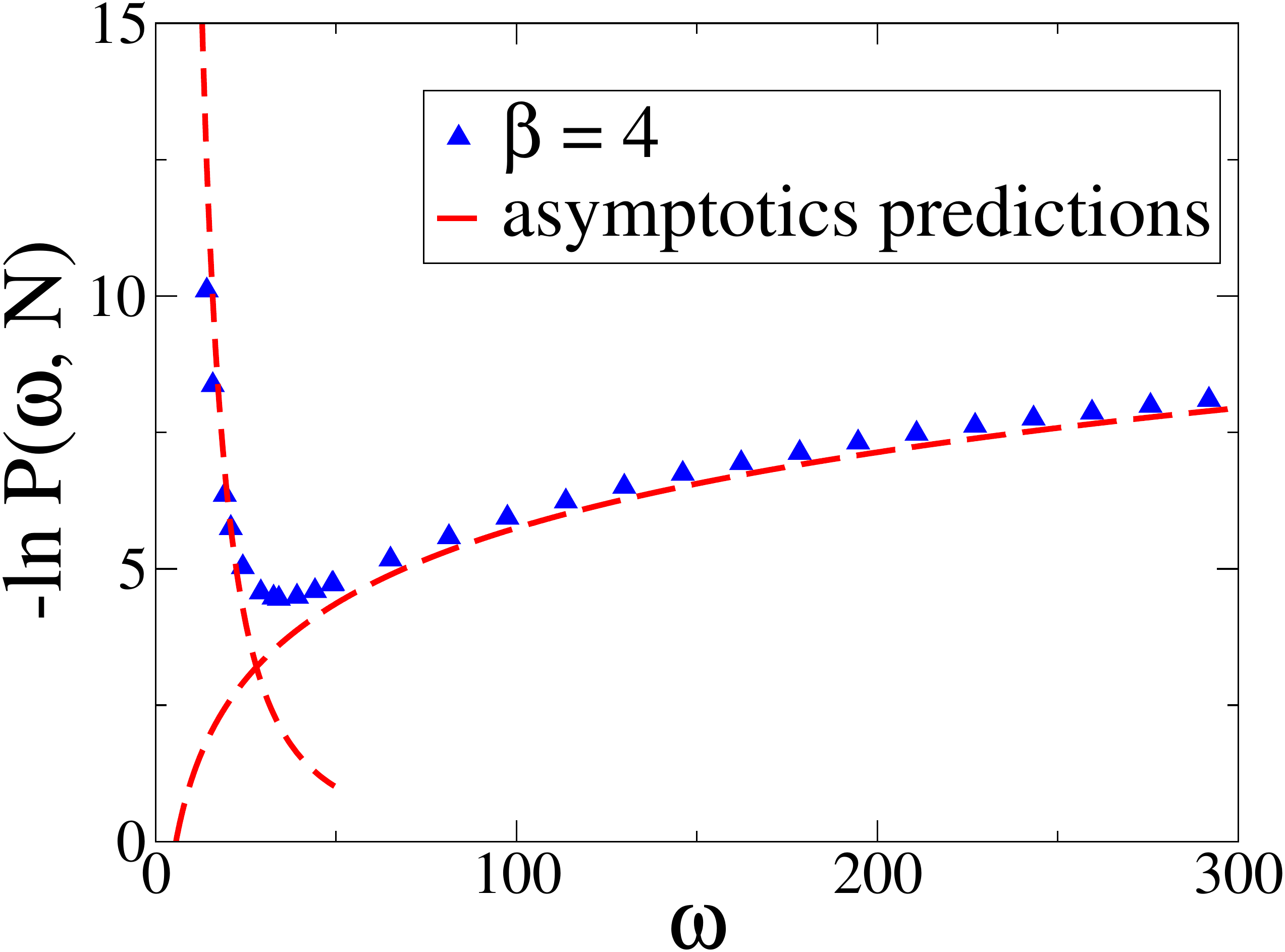}
\end{center}
\vspace*{-0.5cm}
\hspace*{0.5cm}a)\hspace*{4.5cm}b) \hspace*{5cm}c)
\caption{Montecarlo results for the Coulomb gas
    system. The logarithm of the probability density is shown in the
    case $N=100$ for $\beta=1$ in a), $\beta = 2$ in b) and $\beta = 4$ in c). We compare it with our predictions for the left (\ref{large_dev}) and right (\ref{expr_phi}) tails in dashed
    line. For $\beta =2$ we also show the results for $f_2(x)$ (\ref{f2x}), in solid line, which describes the typical fluctuations. Note that one expects that $\psi(w)$ has a support on the full real axis, including $w<0$, where $P(w,N)$ is however extremely small and thus very hard to measure numerically for $N=100$.}\label{fig2}

\end{figure}

{\noindent {\bf Numerical simulations:}} We now present the results of our numerical 
simulations. The distribution of the
   eigenvalues~\eqref{Cauchy_distr} can be directly simulated by
   exploiting the thermodynamical analogy, interpreting the
   eigenvalues $\lambda_{i}$ as a one-dimensional gas of charged
   particles with Coulomb interactions \cite{forrester_book,mehta_book}. Hence the distribution of the maximum eigenvalue can be
   obtained by means of a Metropolis algorithm, the acceptance rate
   between two different configurations being $e^{-\beta \Delta E}$ where
   the energy function $E[\{\mbox{\boldmath$\lambda$}\}]$ is given by
   Eq.~(\ref{energy_coulomb}). Within this numerical scheme, we have tested our analytical predictions in Eqs. (\ref{main_result}), (\ref{asympt}). In Fig.~\ref{fig1} we show a plot of $N P(w,N)$ as a function of the scaled variable $x=w/N$ for $\beta = 2$ and different values of $N = 80, 160, 320$. The good collapse of the different curves is in agreement with the scaling form predicted for the typical fluctuations in the central regime (\ref{main_result}), (\ref{scaling}). The solid line corresponds to a numerical evaluation of $f_2(x)$ in Eq. (\ref{f2x}) obtained by solving numerically the equation for $\tau(x)$ in Eq. (\ref{eq_tau}) (together with the asymptotic behavior $\tau(x) \sim - x/\pi$ for $x \to 0$): the agreement with the numerical data generated by the Metropolis algorithm is excellent. In the inset of Fig. \ref{fig1}, we see that our exact results in Eq. (\ref{asympt}) describe very well the asymptotic behaviors of $f_2(x)$. In addition, in Fig.~\ref{fig2}, we show a plot of $-\ln{P(w,N)}$ as a function of $w$ for $N$ fixed and large ($N = 100$) and for different values of $\beta = 1,2$ and $4$. This plot demonstrates the existence of the three distinct regimes as predicted in Eq. (\ref{main_result}). It shows a very good quantitative agreement with the large deviation functions $\psi(w)$ (\ref{large_dev}) and $\phi(w)$ (\ref{expr_phi}) which we predicted above. Note that $\psi(w)$ has a support on the full real line, not only on the positive axis. However, for $w<0$, $P(w,N)$ is extremely small for $N=100$ and is thus extremely hard to compute numerically (see Fig. \ref{fig2}). Simulations for smaller values of $N$, not shown here, agree with our calculations for $w<0$, although they suffer from strong finite $N$ corrections.     
   
 In summary, we have investigated the fluctuations of the
 largest eigenvalue $\lambda_{\rm max}$ of $N \times N$ random matrices belonging to rotationally invariant 
  Cauchy ensembles. We have considered symmetric ($\beta=1$), Hermitian ($\beta = 2$) and
  quaternionic ($\beta=4$) matrix ensembles. We have identified three different regimes for the pdf of $\lambda_{\max}$: 
  (i) $\lambda_{\max} \ll N$ (left tail), (ii) $\lambda_{\max} \sim {\cal O}(N)$ (central part) and (iii) $\lambda_{\max} \gg N$ (right tail). We have
  obtained exact results for the large deviation tails, both left and right and for any $\beta$, which allows to obtain also the leading asymptotic behaviors
  of the pdf in the central regime, which generalizes the TW distributions known for Gaussian ensembles. These exact results have been
  confirmed by thorough numerical simulations. The exact expression for the central part of the distribution, describing the typical fluctuations of $\lambda_{\max} \sim {\cal O}(N)$, beyond the case $\beta=2$, remains
  a challenging problem.

\textit{Acknowledgments:} SNM and GS acknowledge support by ANR grant
2011-BS04-013-01 WALKMAT. We acknowledge G. Akemann, Z. Burda and M. Tierz for useful discussions. 

\appendix

%\section{Numerical Simulation}

%{\bf [TO BE DONE: DARIO?]}

\end{document}